\documentclass[a4paper,10pt]{llncs}
\usepackage{graphicx}

\usepackage{wrapfig}
\usepackage[english]{babel}
\usepackage{fullpage}

\usepackage{url}  
\usepackage[utf8]{inputenc}  
\usepackage{latexsym}  
\usepackage{amsmath} 
\usepackage{amssymb} 

\usepackage{subfig}

\newcommand{\cpc}[1]{\; \models {\mbox{\small #1}} \Rightarrow \;\;}

\newcommand{\snb}{\mbox{\it SNB}}

\newcommand{\Ne}{\mbox{\em Neg}}

\graphicspath{{figures/}}

\begin{document}

\title{Optimizing Google Shopping Campaigns Structures With Query-Level Matching}

\author{Mathieu Raffinot\thanks{CNRS, LaBRI, University of Bordeaux,
 France. E-mail: {\tt mathieu.raffinot@u-bordeaux.fr}} \and Romain Rivière\thanks{Twenga, Paris, France}}

\institute{}

\maketitle
 
\begin{abstract}
How to bid on a Google shopping account (set of shopping campaigns) with query-level matching like in Google Adwords.\\

\end{abstract}  


\section{Introduction}

In a Google Adwords campaign, one can define the bid entities
(\em{}ie\em{} the criterions) with a set of keywords and associated
matching types (exact, phrase or broad). Those keywords are compared
against the user query, which allows for fine tuning of the user
intent. As in a Google Shopping campaign, the criterions are defined
as the leaves of a product tree defined over the product feed of a
merchant. Fundamentally, in this case, the bidding is done on the
catalog of the merchant, not on the user intent. In this paper, we aim
at taking back control of the user intent with a carefully designed
campaign structure. In particular, we consider a set of keyword of
interest that we want to be able to match to the user query and
associate to a particular set of items in a Google Shopping account
for a specific bid.

\subsection{Structure Of A Google AdWords Campaign for merchants}

\noindent
We will assume, as in every Google Shopping campaign, that the
merchant is able to construct a product feed, which contains the list
of all of its products. Each description of a product has multiple
features, which contains at least a unique identifier itemID, a brand,
and a set of categories. We present below a simple structure of a
Google AdWords campaign instance in the special case of a merchant
campaign.

\begin{enumerate}
\item $\mbox{\tt nike shoes} \cpc{CPC1} \mbox{LandingPage}_1$
\item $\mbox{\tt large tee-shirt} \cpc{CPC2} \mbox{LandingPage}_2,\; \mbox{LandingPage}_3$
\item $\mbox{\tt  garmin chronometer} \cpc{CPC3} \mbox{LandingPage}_4$
\item $\mbox{\tt  adidas running shoes} \cpc{CPC4} \mbox{LandingPage}_5$
\end{enumerate}

In general, the landing pages would contain items that the merchant
want to sell. As a shortcut, we will associate a keyword with a set of
items.

\noindent
\begin{center}
keyword $\cpc{CPC}$ set of itemIDs $I = \{i_1,i_2 \ldots i_k\}$
\end{center}
We will call each of those line a {\em rule}. It means that ideally a given query
{\tt keyword1} should match a product $i_j$ in $I$, bidding at most the
value of Cost Per Clic $(CPC)$ on it.

Let $SK$ the set of all keywords and $SR$ the set of
all rules. The number of such keywords and rules is denoted $n=|SK|=|SR|$,
$n$ can vary from hundreds to thousands of keywords or even
millions. However, active keywords at a given time, meaning those
really leading to conversions during the last days/months, are in general
much less, a few hundreds. Our method is able to handle thousands of
keywords but is generally applied to the few hundreds which convert.

\begin{figure}[htbp]
\centering 
\includegraphics[height=4cm]{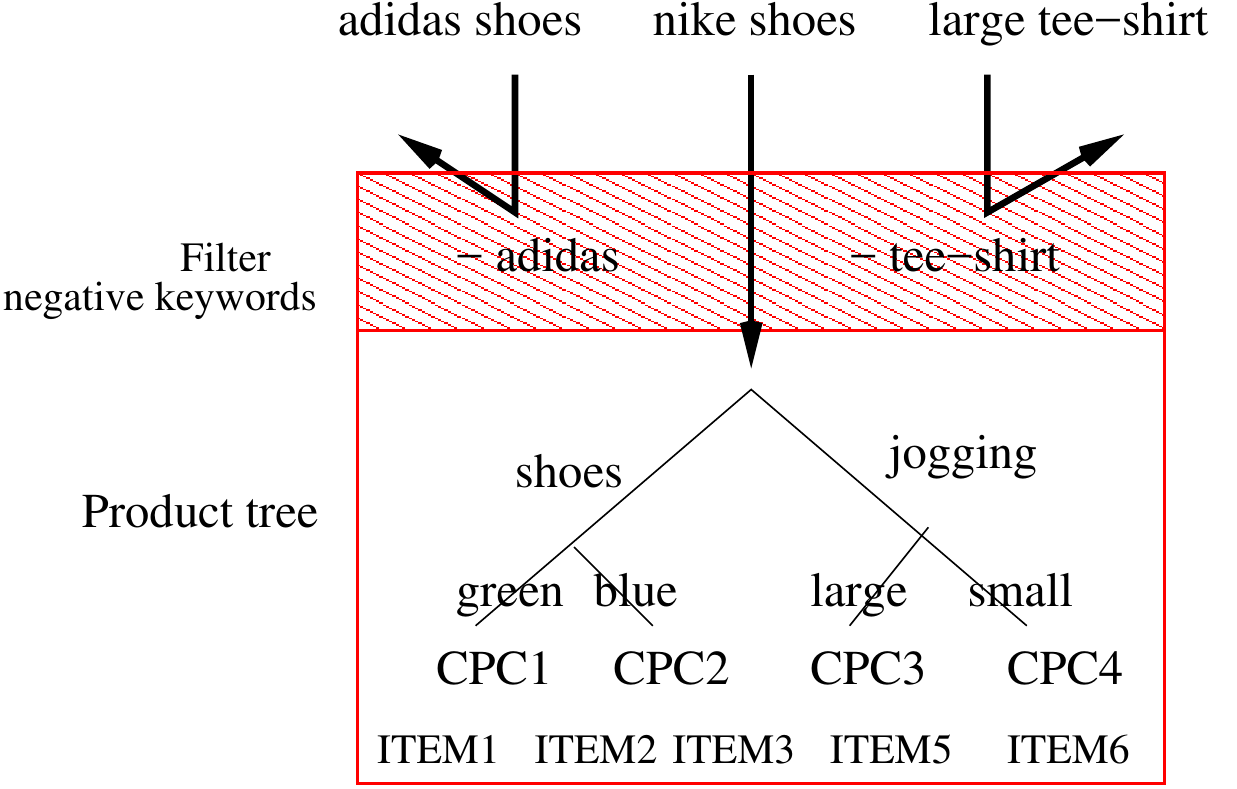}
\caption{an AdGroup of 6 items with a product tree. All negative keywords are of {\em large} type.}
 \label{fig:adgroup}
\vspace{-0.3cm}
\end{figure}

\subsection{Structure Of A Shopping Campaign}

\noindent Each campaign might contain a set of negative keywords. A negative
keyword forbids the entry of the corresponding query keyword in a campaign.
Those keywords are called {\em campaign negative keywords}.

A campaign is structured in AdGroups, each AdGroup corresponding to an item set, placed in
a decision tree (named product tree \-- possibly a single leaf) built using the
features as branches. A leaf of such a tree is named {\bf criterion}.  Each
AdGroup might also contain a set of negative keywords. Figure
\ref{fig:adgroup} represents an instance of an AdGroup with a product
tree.

Let $p$ be a keyword, we denote $w(p)$ the set of continuous subwords
that $p$ contains. For instance, $w(\mbox{\tt nike large shoes}) =
\{\mbox{\tt nike},$ $\mbox{\tt large},$ $\mbox{\tt shoes},$ $\mbox{\tt
  nike shoes},$ $\mbox{\tt large shoes},$ $\mbox{\tt nike large
  shoes}\}.$ We denote $s(p)$ the set of words composing $p$.

The negative keywords of a campaign or an AdGroup can be of three
types: {\em exact}, {\em large} or {\em phrase}, which differ on the type
of matching against the query. An {\em exact} negative keyword matches a
query which is exactly the keyword. A {\em phrase} negative keyword $p$
matches any query $q$ such that $p \in w(q)$.  A {\em large} negative
keyword $p$ matches any query $q$ such that $s(p) \in s(q)$.

A shopping campaign can be conceived as a jar which lid is a filter composed of
negative keywords.  A query keyword tries to enter the jar, but is stopped by
the filter if one negative keywords matches, in an exact, phrase or large match.
If not the case, the keyword goes through the filter, enters the jar, and faces
the AdGroups. An AdGroup can also be seen the same way, a jar with a filter, but
to which is associated a product tree.

The limit on the number of keyword of a campaign or an AdGroup is denoted $L$
and is at this time around $20\,000.$ One problem which intrinsically complicates
our approach below is that $L$ might be less than $n$. In this version of our
article, we consider that $n<L$, which is, as stated above, quite always the
case when considering rules leading to conversions. This implies that all
keywords would fit as negative in either campaigns or AdGroup negative keywords
sets.

\begin{figure}[htb!] \centering
\includegraphics[height=5cm]{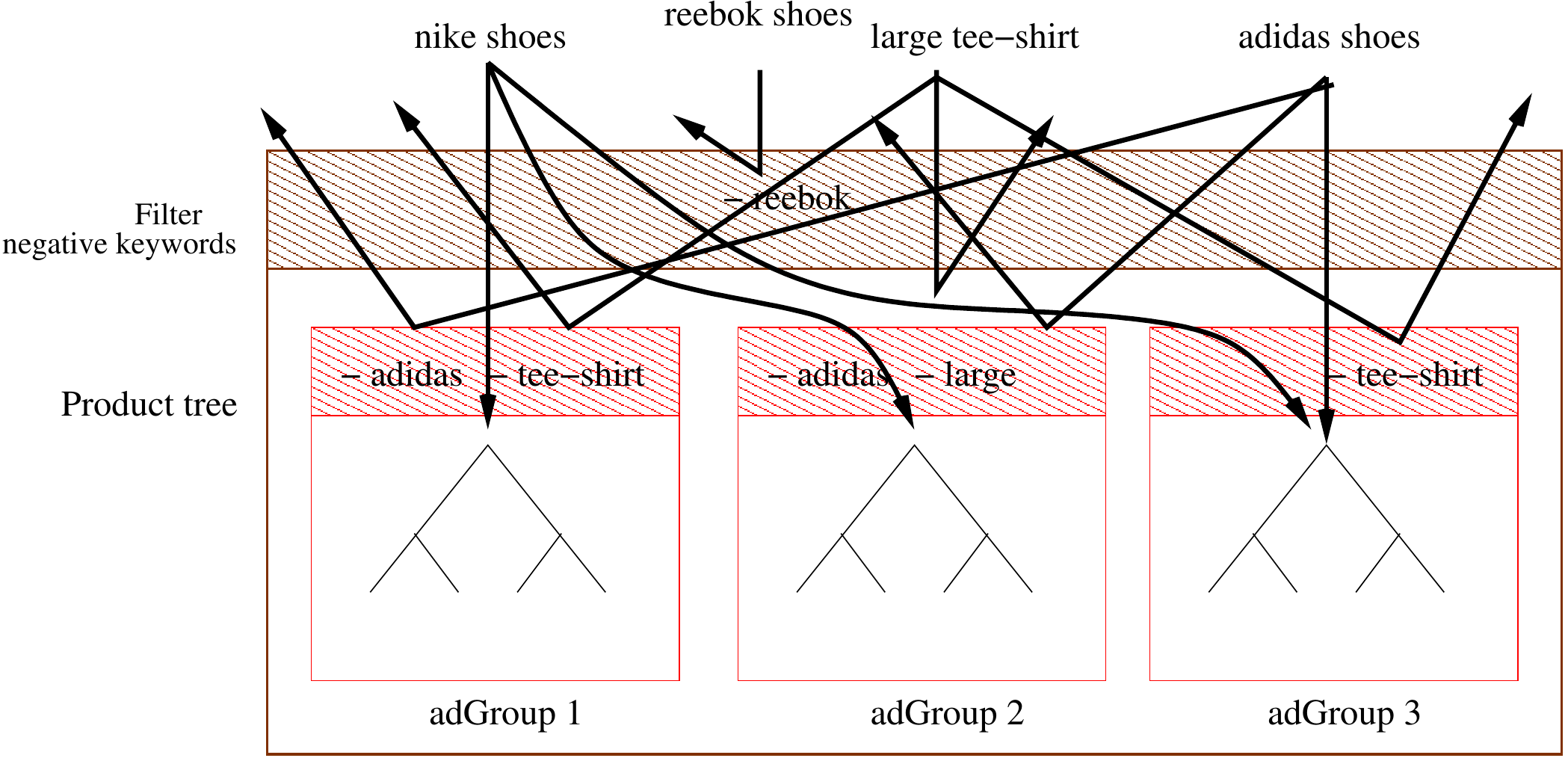} \caption{A shopping campaign
containing 3 AdGroups with some keyword trajectories. All negative keywords are of {\em large} type.}
 \label{fig:threeadgroup}
\vspace{-0.3cm}
\end{figure}

The way a query keyword passes through campaigns and AdGroups is named its {\em
trajectory}. Figure \ref{fig:threeadgroup} shows some trajectories.

\subsection{Structure Of A Shopping Account}

\begin{figure}[htb!]
\centering 
\includegraphics[height=3.5cm]{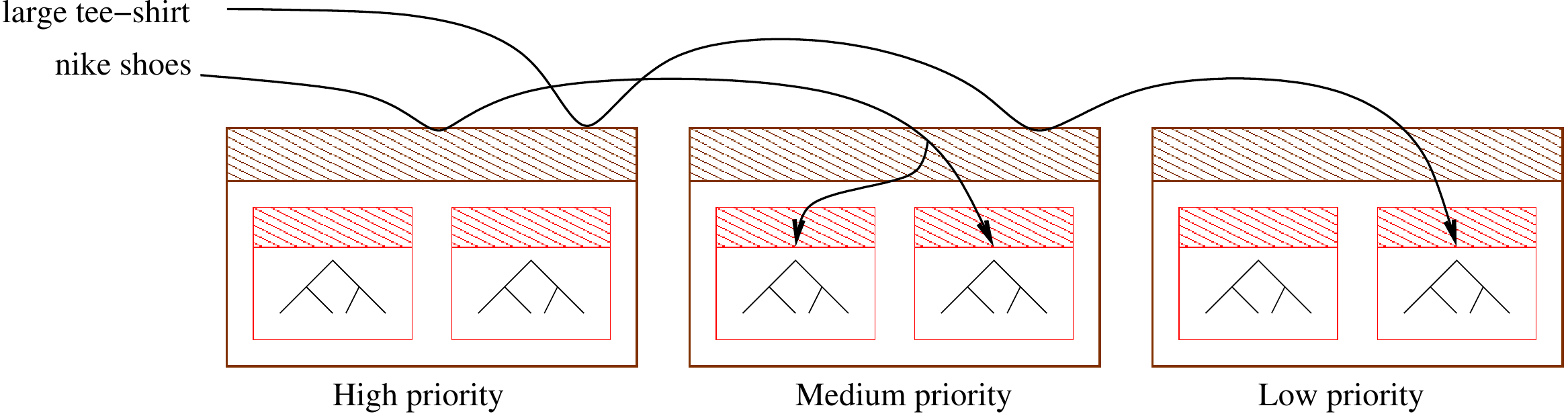}
\caption{A shopping account containing 3 campaigns of high, medium and
  low priorities.}
 \label{fig:account}
\vspace{-0.3cm}
\end{figure}

A shopping account may contains many campaigns, that are classified by
priority. There are 3 priority levels: high, medium and low. Given a
query keyword, a match is attempted in every campaign by decreasing
priority order. Once a match found, the remaining campaigns are not
any more checked. Figure \ref{fig:account} shows such an account with
some query keywords trajectories.

\subsection{Motivation And Objectives}

The main drawback of using google shopping is that the matching
between the query keyword and the items is decided by Google with very
low control to the user. This is a deep difference with AdWords
campaigns. This induces many difficulties for digital advertising
agencies to optimize the shopping account and to model the bids.

However, some high-tech agencies recently proposed new shopping account
structures using the campaign priorities and/or the keywords to better control
the matching. This requires to better control the trajectories of queries
through the account, whatever the query.\\

\noindent
To our knowledge, two main approaches have been designed up to far:
\begin{enumerate}
\item use campaign level priorities to distinguish between general, long tail,
  and specific requests \cite{KW2015,MR2014}.
\item use AdWords rules and campaign negative keywords to design 2
  levels campaign account guiding query keyword trajectories to land in a
  specific AdGroup \cite{DG2015}.\\
\end{enumerate}

\noindent
In this article we mix and extend both approaches to:
\begin{enumerate}
\item use the 3 campaign priorities to distinguish general, less general but branded, and 
eventually query keywords that has been chosen in a pre-defined AdWord campaign
\item use AdWords rules and campaign negative keywords to design 3
  levels campaign account guiding query keyword trajectories to the set of
  item(s) specified by the rule its belongs to.
\item minimize the number of negative keywords, of campaigns and
  AdGroups, to reach this goal.
\end{enumerate}
This structure permits to specify a CPC for each rule, thus getting
the lowest possible granularity as a pre-requise for a smart bidding
strategy.\\

This article is structured as follows: in Section \ref{sec:structure} we first
explicit the structure of a campaign account we propose and prove that it
corresponds to what we expect. We bound in Section \ref{sec:bound} the number of
keywords required. We then explain in Section \ref{sec:lowering} how to strongly
lower on average the negative keyword number using a keyword reduction based on
an ad-hoc heuristic. We explain in Section \ref{sec:updates} how to perform
several types of update on this structure.  Eventually, in Section
\ref{sec:example} we present a complete example.

Let us denote $SB$ the set of brand names the merchant sells, and
$\snb$ the set of brand names that the merchant does not sell but are
known in the retail domain of the merchant. For instance, a merchant can
sell the brands {\tt nike} and {\tt adidas} but not {\tt reebok}. Then
$SB=\{\mbox{\tt nike},\mbox{\tt adidas},\mbox{\tt garmin}\}$ and $\snb=\{\mbox{\tt
  reebok}\}.$ For each brand $b\in SB$ we assume a product tree $PT(b)$
to be given. For instance for the brand {\tt nike}, a $PT(\mbox{\tt
  nike})$ could be $PT(\mbox{\tt nike})=[[\mbox{ \tt shoes}, 
    CPC12]$ $[\mbox{\tt jogging}, CPC14]$ $[others, CPC15]]$, where {\tt
  shoes} and {\tt jogging} are two pre-defined categories.

\section{Our Structure Of Shopping Account}
\label{sec:structure}


We propose the following 3 levels structure. The first level is a high
priority campaign which filters queries that are the more general, non
branded and do not belong to $SK$. The medium level is again a single
campaign, but which does not filter branded keywords that are not in
$SK$. The third level consists of a series of campaigns which permit to
exactly match the keywords in $SK.$

\subsection{High-priority Campaign $C_1$}

\begin{figure}[htb!]
\centering 
\includegraphics[height=3.3cm]{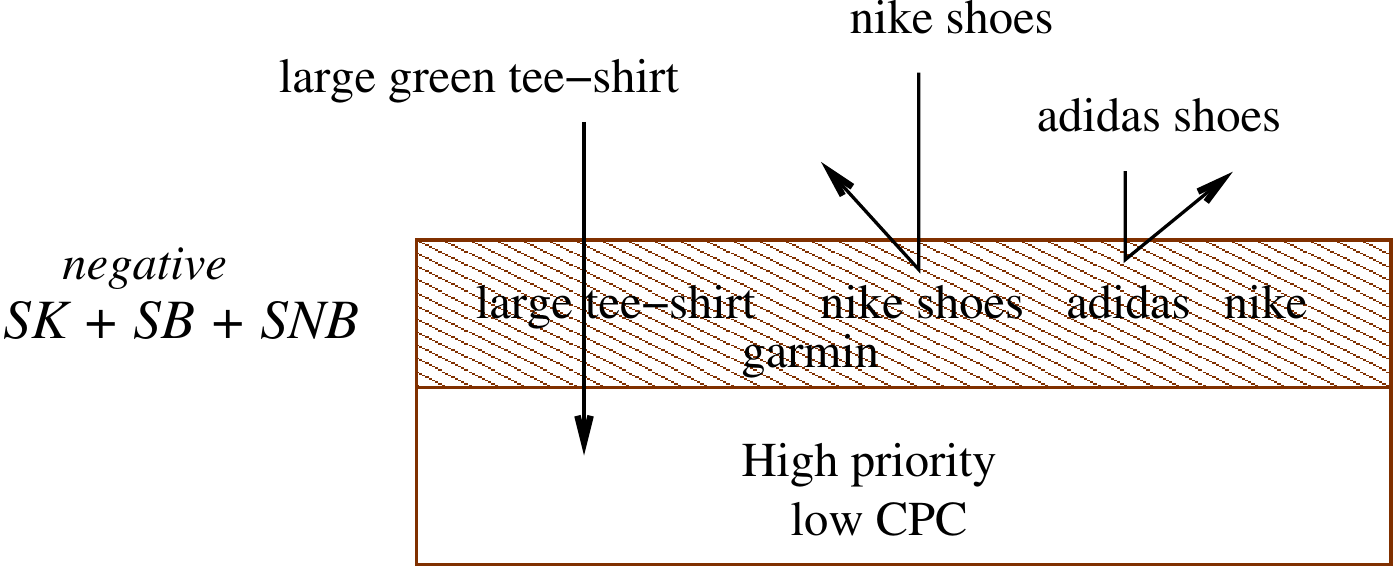}
\caption{High-priority campaign. Negative keywords are $SK(exact)$ $\cup SB(phrase)$ $\cup SNB(phrase)$,
  filtering all the pre-defined keywords and also keywords not in $SK$
  but containing brand names (of the site or not).}
 \label{high}
\vspace{-0.3cm}
\end{figure}

We create one high priority campaign $C_1$ which role is to filter all {\em a
priori interesting} keywords, that is, the keywords which are stopped by $C_1$
are $SK (exact) \cup SB (phrase) \cup \snb (phrase)$. When one keyword passes the
filter, it is matched by Google to possibly a set of items on which we have no
control. Thus the idea is to fix low value CPCs at this point, but to keep
watching those ``refereers'' which could be integrated latter to the structure
if they truly convert. Figure \ref{high} shows such an high level campaign.

\subsection{Medium-priority Campaign $C_2$}

\begin{figure}[htb!]
\centering 
\includegraphics[height=6cm]{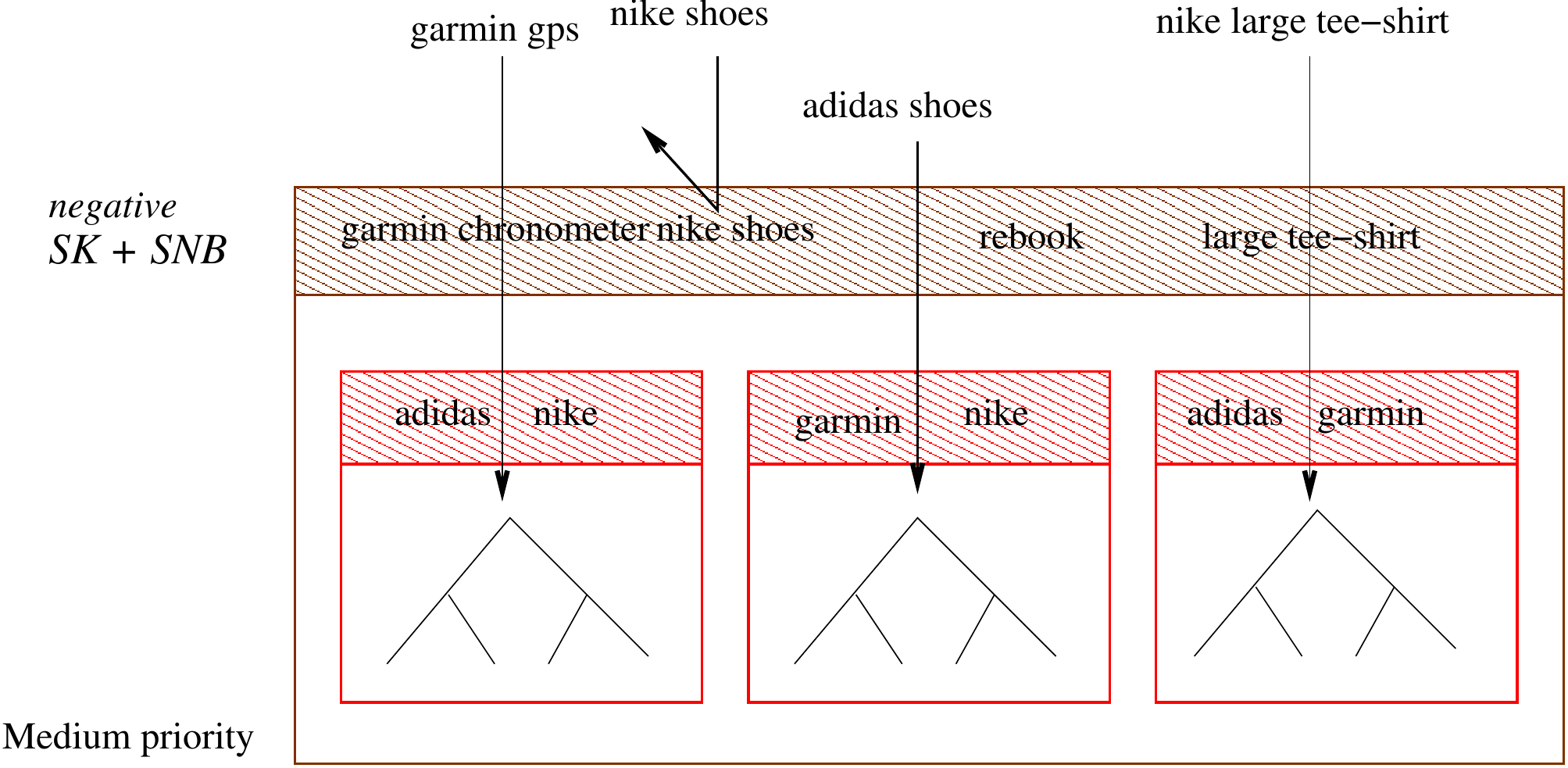}
\caption{Medium-priority campaign. Negative keywords are $SK(exact)$ $\cup \snb(phrase)$,
  filtering all the pre-defined keywords. However, keywords not in
  $SK$ but containing brand names in $SB$ pass the negative filter at
  this campaign level. One AdGroup by brand name.}
 \label{medium}
\vspace{-0.3cm}
\end{figure}

The medium-priority campaign $C_2$ (fig. \ref{medium}) is devoted to
filter branded keywords that are not in $SK$. Then the CPC are
adjusted by brand name. The negative of the campaign are $SK$ and
$\snb$ (but not $SB$). The campaign contains one AdGroup per brand name,
say $b_i \in SB$. The AdGroup corresponding to $b_i$ has $SB\setminus b_i$ 
for negative keywords. This way, only queries containing $b_i$
pass the filter. The AdGroup may contain a product group allowing to
better adjust CPCs with the brand features.

\subsection{Low Priority Campaigns $C^i_3$}

\begin{figure}[htb!]
\centering 
\includegraphics[width=15cm]{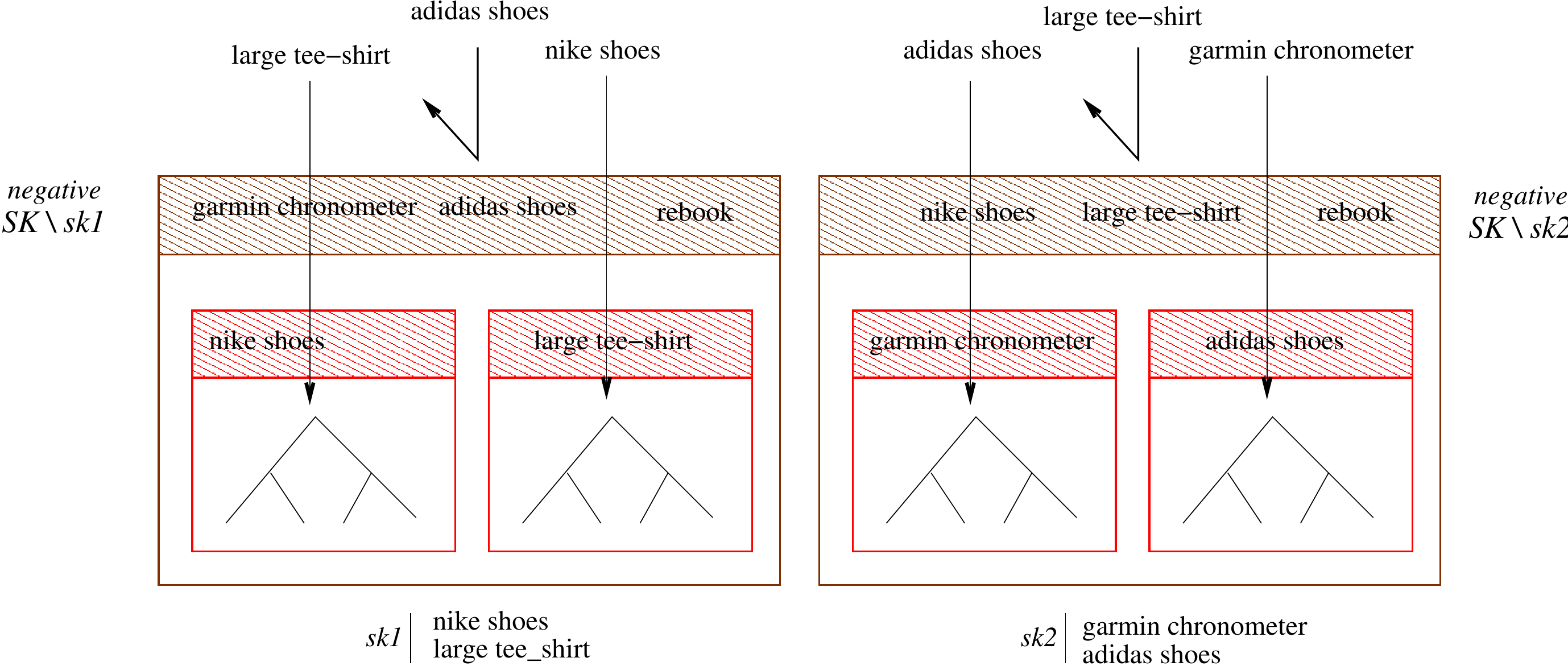}
\caption{One low-priority campaign (among $lp$ such campaigns)
  corresponding to the set $sk$ of keywords. Campaign negative keywords are
  $(SK\setminus sk) \cup \snb$. The campaign contains $sk$ AdGroups, one
  for eack keyword in $sk$. The AdGroup corresponding to $k_i \in sk$
  has $sk\setminus k_i$ as negative keywords.}
 \label{low}
\vspace{-0.3cm}
\end{figure}

The set of keywords $SK$ is divided in $k$ groups $sk_1, sk_2, \ldots
sk_k$. We create $k$ campaigns $C^i_3, 1 \leq i \leq k,$ each $C^i_3$
being built to match all keywords in $sk_i$ and only them, and thus has
for negative keywords $SK \setminus sk_i \cup \snb.$ Then, one AdGroup
is created for each keyword $l \in sk_i, 1 \leq l \leq |sk_i|$,
filtering all the other keywords in $sk_i$. Thus the AdGroup
corresponding to $j$ has $sk_i \setminus j$ for negative
keywords. Figure \ref{low} shows such a series of campaigns on our
continuing example, where $k=2$ and $sk_1=\{${\tt nike shoes}, {\tt large
tee-shirt}$\}$, $sk_2=\{${\tt garmin chronometer}, {\tt adidas shoes}$\}.$

\subsection{Structure Main Property}

We prove the three main properties of our structure. Let $q$ be a query keyword. 

\begin{property}
If $q \in SK,$ there exist a unique $C_3^j$ such that $q$ is
associated with a unique AdGoup of $C_3^j.$
\end{property}
\begin{proof}
$q$ does not enter $C_1$ nor $C_2$ since it belongs to $Neg(C_1)$ nor
  $Neg(C_2).$ $q$ is then stopped by all but one low priority
  campaign, say $C_3^k$, since it belongs to a unique set $sk_k \in
  SK$ and that all others low priority campaigns stop $sk_k$. In
  $C_3^k$, $q$ passes the filter of a unique AdGroup corresponding to
  the keyword $q$ itself.
\end{proof}

The following property states the case when a query does not match any
keyword in $\in SK$ but contains a brand name.

\begin{property}
If $q \not\in SK$ but there exits a unique $w \in w(q)$
such that $w \in SB$, $q$ is associated to a unique AdGroup in $C_2$.
\end{property}
\begin{proof}
As $SB \in Neg(C_1)$ as {\em phrase}, $q$ does not enter $C_1$. Then $q$
is tested against $C_2$. As $q \not\in SK,$ and that $SB \not\in
Neg(C_2),$ $q$ enters $C_2$. Then $q$ is blocked by all AdGroups
containing $w$ as {\em phrase}. Only one AdGroup (by construction) does not
stop $w$, and $q$ is thus associated to this AdGroup.
\end{proof}

The following property states the case when a query does not match any
keyword and does not contain a brand name.

\begin{property}
If $q \not\in SK$ and $w(q) \cap SB = \emptyset,$ $q$ is recognized in $C_1$.
\end{property}
\begin{proof}
$q$ is not stopped by any negative keyword of $C_1$, thus $q$ passes through the filter of $C_1$.
\end{proof}

\section{Bound On The Number Of Negative Keywords}
\label{sec:bound}

Our structure permits fine-grained bidding on queries by assigning to
each rule a specific AdGroup which is only reachable by the
keyword of the rule. Let us now count and optimize the number of negative keywords
it requires. We recall that $n = |SK|$. We denote $m=|SB|$ and $m'
=|\snb|$ and that $k$ is the number of splits of $SK.$

\subsection{Number Of Negative Keywords}

\begin{enumerate}
\item 1 high priority campaign, $n+m+m'$ campaign negative keywords.  
\item 1 medium priority campaign, $n+m'$ campaign negative keywords.
\begin{enumerate}
\item $m$ AdGroup with $m-1$  negative keywords each
\end{enumerate}
\end{enumerate}
Thus the number of negative keywords for the high and medium priority
campaign is $2n+2m'+m+(m-1)m =2n+2m'+m^{2}.$\\

Let us count now the negative keywords of low level campaigns. Each
$sk_i$ set leads to a campaign $C_3^i$ which contains $\sum_{l\neq i,
  1 \leq l \leq k} |sk_l|+ m'.$ The total over all $C_3^i$ is then
$$\sum_{1 \leq i \leq k} \left( \sum_{l\neq i, 1 \leq l \leq k} |sk_l|
+ m' \right) = (k-1) \sum_{1 \leq l \leq k} |sk_l| = (k-1) n + k m'$$
campaign negative keywords.\\

Each campaign $C_3^i$ also contains $|sk_i|$ AdGroups, each such
AdGroup containing $|sk_i|-1$ AdGroup negative keywords, thus
$|sk_i|*(|sk_i|-1) = |sk_i|^2 -|sk_i|$ keywords. Summing over all
$C_3^i$, this leads to $\sum _{1 \leq i \leq k} |sk_i|^2 -|sk_i| =
\sum _{1 \leq i \leq k} |sk_i|^2 \;-n $

The whole total number of negative keywords is thus $NK = 2n+2m'+m^{2} +
(k-1) n + \sum _{1 \leq i \leq k} |sk_i|^2 \;-n = m^2 + (k+2) m'+ kn +
\sum _{1 \leq i \leq k} |sk_i|^2.$

\subsection{Worst Case Optimization}

We want to minimize $NK$. Because of the square in the last term. This
appears when the $sk_i$ are all of the same cardinal of $n/k$ keywords,
leading to $NK = m^2 + (k+2) m'+ kn + k (\frac{n}{k})^2.$ Considering
that $m$ and $m'$ are small compared to $n$, the minimum is reached
when $k$ is close to $\sqrt n $. This leads to a number of keywords of
$NK = m^2+(\sqrt n +2) m' + 2 \sqrt n \, n.$ We show below some
values on real data to visualize the number of negative keywords this
approach requires.

\begin{center}
\begin{tabular}{|l|r|r|r|r|}
\hline
Name & $SK$ & m & m' & $NK$\\
\hline
Site1 & 3000  & 100 & 30 &  340337\\
Site2 & 7000 &   1  &  0 &  1171324\\
Site3 & 10000 & 30 & 20 &  2002940 \\
Site4 & 10000 & 1000 & 40 & 3002040\\
\hline
\end{tabular}
\end{center}

\section{Lowering The Number Of Negative Keywords}
\label{sec:lowering}

\begin{figure}[b]
\centering 
\includegraphics[width=8cm]{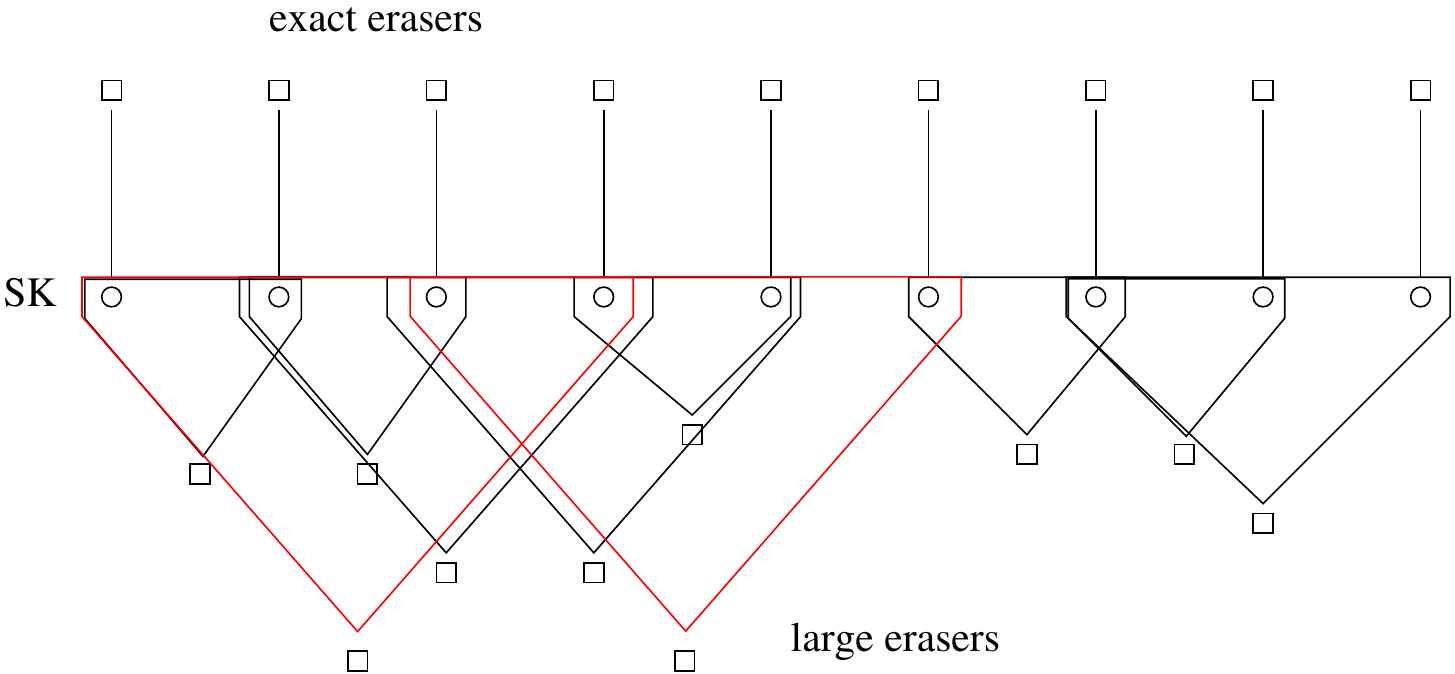}
\caption{Exact and large erasers of a set of keywords $SK$. A keyword
  is drawn as a circle while an eraser as a black square.}
 \label{erasers1}
\vspace{-0.3cm}
\end{figure}

The total worst case number of negative keywords is high compared to
the number of $NK$. We propose now an heuristic to reduce on average this
number of keywords. This technique is based on a new notion of
``eraser'' that we define formaly. 

\begin{definition}
A {\em large eraser} of a set $P=\{p_1,\ldots,p_j\}$ of keywords is a
set of words $e$ such that $e \in s(p_i).$ 
\end{definition}

\begin{definition}
An {\em exact eraser} of a set $P=\{p\}$ of a single keyword is 
the keyword $p$ itself.
\end{definition}

The set of erasers (large or exact) of $P$ is denoted $E(P).$ The {\em image}
of a given eraser $E$ relatively to a set $P$ is the set of
keyword(s) of $P$ it erases. Its size is denoted $I_P(E).$ Note that
if $E$ is an exact eraser on $P$, $I_P(E)=1.$ When the set of keywords
$P$ is non-ambiguous, $I_P(E)$ is simplified as $I(E).$

Usually the set $P$ is contained in a larger set, say $SP$, and we
introduce the notion of {\em strict eraser}.

\begin{definition}
Let $SP = \{p_1,\ldots,p_j\}$ and $P\subsetneq SP$. A {\em strict
  eraser} of $P$ relatively to $SP$ is an eraser of $P$ which is not
an eraser of any $p \in SP\setminus P.$ The set of strict erasers of
$P$ relatively to $SP$ is named $E_{SP}(P).$
\end{definition}

Note that an exact eraser is always strict. Let $P(SK)$ be the set of
all subsets of the set of keywords $SK$. We consider the set $ER(SK)$
of all erasers of all sets in $P(SK).$

The idea to limit the set of negative keywords is to replace subsets of
negative keywords of a given $C_3^i$ campaign by a set of smaller
cardinal of large erasers, but which are not erasers of the $sk_i$
keywords that $C_3^i$ must not filter. Those erasers, as they are
large, may filter many more keywords that would have been filtered by
the initial list. But this is no issue since the additional
keywords erased will either be accepted by another $C_3^j$ campaign,
either be accepted by a higher priority campaign $C_2$ or $C_1.$

\begin{figure}[htb!]
  \centering
  \subfloat[Filtering erasers which image size $\leq \sqrt n = 3$]
           {\label{fig:erasers2-a}\includegraphics[width=0.45\linewidth]{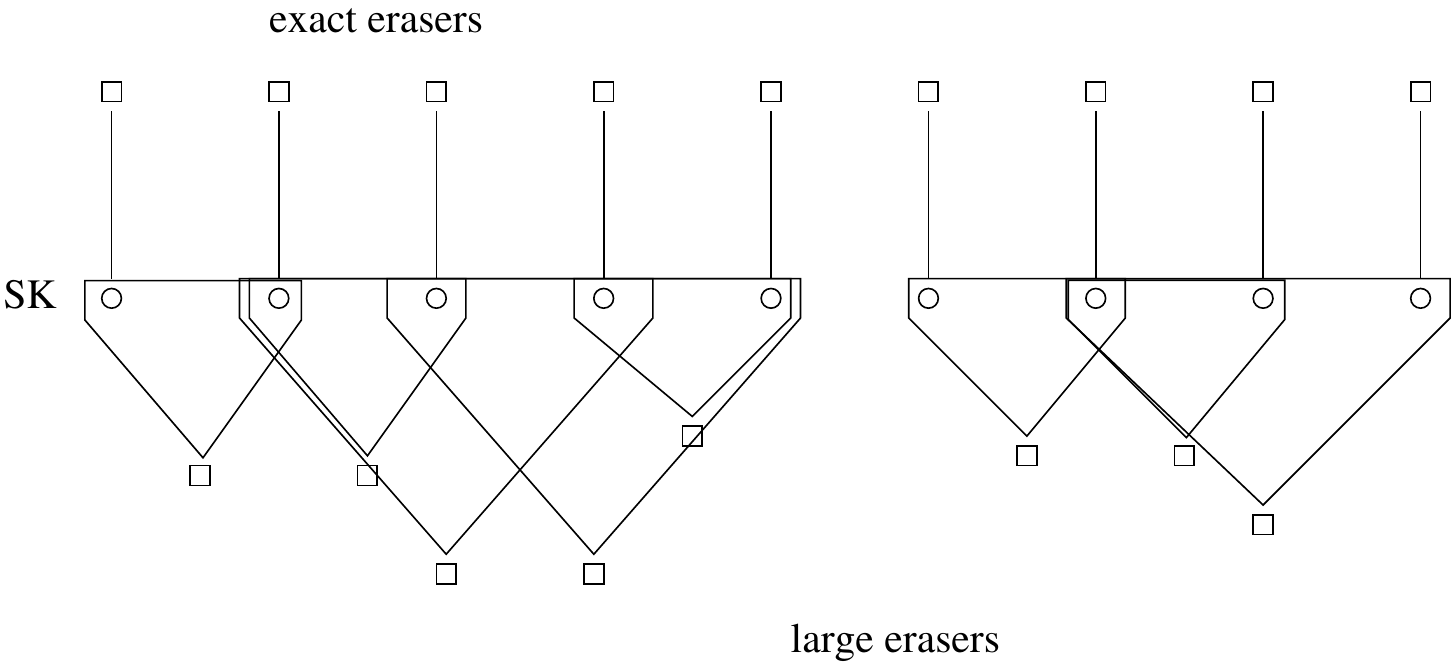}} \quad
  \subfloat[Select as few erasers as possible to cover $SK$]{\label{fig:erasers2-b}\includegraphics[width=0.45\linewidth]{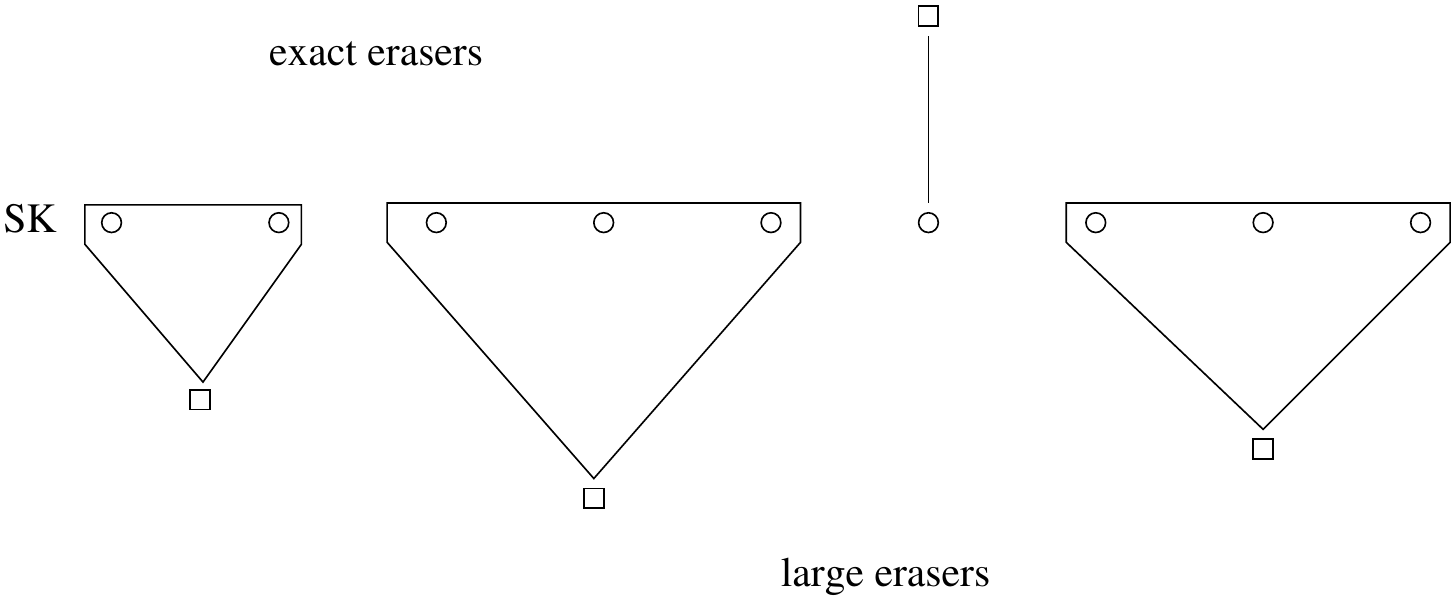}}

  \subfloat[Grouping remaining erasers together]{\label{fig:erasers2-c}\includegraphics[width=0.6\linewidth]{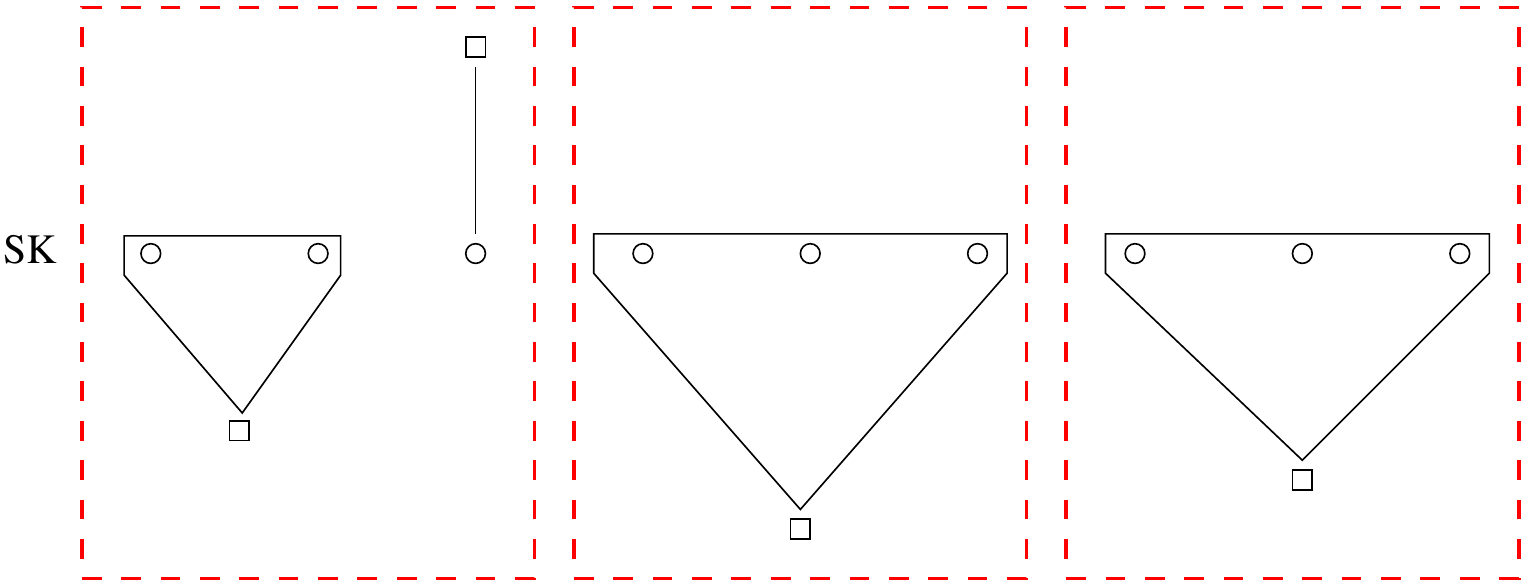}}
  \caption{Scheme of our approach to balance the set of keywords in
    about $\sqrt n = 3$ groups of size of more or less $\sqrt n = 3$
    keywords using the minimum of large or exact erasers.}
\end{figure}

The problem becomes informally to balance the set of keyword in about
$\sqrt n$ groups of size of more or less $\sqrt n$ keywords using the
minimum of large or exact erasers. Figure \ref{erasers1} shows the
initial state. One approach we tested to reach this goal is to: 

\begin{enumerate}
\item filter erasers by image size and only keep those which
  image is less than or equal to $\sqrt n$ (see Figure \ref{fig:erasers2-a}). 
\item select as few erasers as possible to cover $SK$ with non-overlapping images (see Figure \ref{fig:erasers2-b}).
\item group remaining erasers together to merge their images to form larger
  images but still of size less than or equal to $\sqrt n$ (see Figure \ref{fig:erasers2-c}). 
\end{enumerate}

Step 2 is a classical problem called {\rm exact weighted set packing}
which is NP hard. However, there does not exist a standard guarantied
heuristic to approximate it.

We propose an heuristic based on a weighted graph coloring. We define
the graph $GE$ as follows: its vertices are the set of all large erasers of
image sizes less than $\sqrt n$.  Let $E_1$ and $E_2$
be two erasers. There is an arc $(E_1,E_2)$ in $GE$ if the images of
$E_1$ and $E_2$ intersect. The two erasers are then said incompatible.
Each eraser node $E$ is weighted by $I(E)$.

We use Welsh-Powell heuristic \cite{Welsh85} which returns a color for each
node $E$, denoted $c(E)$. For each such color, we sum all images of all nodes colored the same
and we select the color $c$ leading to the maximum such sum. We
then compute the union of all images of all eraser nodes colored $c$:
$$\mbox{SC} = \cup_{E | c(E)=c} I(E)~.$$ Eventually, for each keyword
in $P\setminus \mbox{SC},$ we add the keyword as its own exact
eraser. The remaining set of erasers \mbox{SE} is thus:

$$\mbox{SE}= \left \lbrace \begin{array}{l} E \;|\; c(E)=c\;
  (large)\\ \mbox{SC}\setminus P \; (exact) \end{array} \right.$$

We define two algorithms: $\mbox{\em reduce}(S)$ where $S$ is a set of
keywords, which returns a set of erasers $E_{S}$, large or exact of
$S$. We also define $\mbox{\em expand}(E_{S})$ which returns the
original set $S$.

\subsection{Experimental Results}

We performed some tests on real data for 2 marchant sites, with 3000 rules
for the first site and 7000 for the second. The following table shows
our results, in which we also exhibit the number of erasers (neras) in our
intersection graph and its number of transitions (ntrans).

\begin{center}
\begin{tabular}{||l|r|r|r||r||r|r|r|r||}
\hline
Name & $SK$ & m & m' & $NK$ & neras & ntrans & h = heuristic & h/$NK$\\
\hline
Site2 & 1000 &   1  &  0 &  63246  & 22332 &  908790 &  {\bf 18193} & 0.29\\
Site2 & 2000 &   1  &  0 &  178886 & 39732 & 1705080 & {\bf 44215}  & 0.25 \\
Site2 & 3000 &   1  &  0 &  328634 & 56989 & 2530764 & {\bf 85639}  & 0.26\\
Site2 & 4000 &   1  &  0 &  505965 & 71969 & 3249978 & {\bf 127334} & 0.25\\
Site2 & 5000 &   1  &  0 &  707107  & 84299 & 3856870 & {\bf 169946} &  0.24\\
Site2 & 6000 &   1  &  0 &  929516  &  96588 &  4424679 & {\bf 227781} &  0.24\\
Site2 & 7000 &   1  &  0 &  1171324 &  107993 & 5033879 & {\bf 290702} &  0.25\\
\hline
Site1 & 1000  & 100 & 30 &  74254 & 2653 & 23110 & {\bf 20993} &  0.28 \\
Site1 & 2000  & 100 & 30 &  190287 & 4763 & 42748  & {\bf 56155 } & 0.29\\
Site1 & 3000  & 100 & 30 &  340337 &7123 & 68661 & {\bf 118187} & 0.35 \\
\hline
\end{tabular}
\end{center}

Our tests remain succint, and we observe roughly a reduction (column
h/neras) of about $\frac{2}{3}$ of the number of negative keywords required
to organise the shopping account. We plan however to develop more
tests and study more parameters, like the size of the graph of eraser
intersections compared to the distinct word number and the maximal
number of words in a keyword in the entry set.

\section{Updates}
\label{sec:updates}

The structure must allow to optionally update the merchant stream
easily. More specifically, a user must be able to:
\begin{enumerate}
  \item[op1:] add a rule $R^{a} : \mbox{\tt key-add} \cpc{CPC1} I = \{\mbox{\em Item}_1,\mbox{\em
      Item}_2, \ldots ,\mbox{\em Item}_k \}$
  \item[op2:] remove a rule on existing items $R^{e} : \mbox{\tt key-rm} \cpc{CPC2} I = \{ \mbox{\em Item}_1,\mbox{\em
    Item}_2, \ldots ,\mbox{\em Item}_k \}$
  \item[op3:] remove an item $\mbox{\em Item-rm}$ and remove/slit all rules associated with it.
    
\end{enumerate}

The idea is to update smoothly the structure, touch as few AdGroup as
possible, until the campaign becomes too unbalanced. Only then a large
update is performed.

\subsection{op1: Add A Rule}
\label{adding}

There exist many possible strategies to add a rule on existing items,
with possibly distinct objectives. For instance, one objective is
possibly that the structure remains meaningful for an account
manager, or a client. Another objective can be to minimize the number
of AdGroups touched. A third objective is to globally minimize the
number of negative keywords. We propose an algorithm for this last
goal below. Two cases might occur:

\begin{enumerate}
\item there exists a set $\{i_1 \ldots i_k\}$ such that $ 1 \leq j
  \leq k, \; \Ne(C_3^{i_j})$ does not erase $\mbox{\tt key-add}.$ Then
  let $ 1 \leq l\leq k, $ be the indice such that $|sk_{i_l}|$ is
  minimal. Then, (a) let $sk_{i_l} \leftarrow sk_{i_l} \cup \{
  \mbox{\tt key-add}\},$ (b) for all $1 \leq j\leq k, j \not= l,
  Neg(C_3^{i_j}) \leftarrow Neg(C_3^{i_j}) \cup \mbox{\tt key-add}.$
  Then, in $C_3^{i_l}$, (c) for each AdGroup {\em ad} in $C_3^{i_l}$,
  $\Ne(\mbox{\em ad}) \leftarrow \mbox{\tt key-add},$ and eventually
  we create in $C_3^{i_l}$ a new AdGroup $\mbox{\em adnew}$ and set
  $\Ne(\mbox{\em adnew}) \leftarrow sk_{i_l} \setminus \mbox{\tt key-add}.$
\item all $\Ne(C_3^{i})$ erase $\mbox{\tt key-add}.$ There are two main possible strategies:
  \begin{enumerate}
  \item either modify the negative set of one of the $C_3$ campaign
    not to filter $\mbox{\tt key-add}$ anymore, and then apply point 1. We discuss this approach below.
  \item either create a new $C_3^h$ campaign, setting $\Ne(C_3^h)
    \leftarrow SK.$ In this campaign we create a single AdGroup
    $\mbox{\em adnew}.$
\end{enumerate}
\end{enumerate}

Point 2-(a) requires a specific algorithm. After the reduction of the
number of negatives using the heuristic of Section \ref{sec:lowering},
large erasers permit to lower the number of exact negatives, but exact
negatives can always remain a last option if erasers are too large and
erase the new keyword $\mbox{\tt key-add}$ we need to add. Thus,
the approach is to identify which group $sk_l \cup \mbox{\tt key-add}$
leads to the minimum increase of the negative erasers of all $C_3^i$.

\subsection{op2: Remove A Rule On Existing Items}

Removing a rule is not difficult, roughly it suffices to remove $\mbox{\tt
  key-rm}$ for each negative set of all campaigns it belongs to.

\subsection{op3: Remove An Item With Rules Associated With It}

To remove a specific item, it suffices to remove each rule where the
item is the only target of the rule.


\section{Larger Example Of $C_3$ Campaigns}
\label{sec:example}

\noindent
{\footnotesize
\begin{tabular}{l|l}
$\mbox{\tt nike shoes} \cpc{CP1} \mbox{Item}_1$
& $\mbox{\tt large tee-shirt} \cpc{CPC2} \mbox{Item}_2,\; \mbox{Item}_3$ \\
 $\mbox{\tt  garmin chronometer} \cpc{CPC3} \mbox{Item}_4$
& $\mbox{\tt  adidas running shoes} \cpc{CPC4} \mbox{Item}_5$ \\
 $\mbox{\tt nike soccer white} \cpc{CP5} \mbox{Item}_1$
& $\mbox{\tt soccer colored mens} \cpc{CPC6} \mbox{Item}_1$ \\
 $\mbox{\tt adidas superstar} \cpc{CPC7} \mbox{Item}_5$
& $\mbox{\tt adidas superstar sneaker} \cpc{CPC8} \mbox{Item}_5$ \\
 $\mbox{\tt  large superstar shoes} \cpc{CPC9} \mbox{Item}_2$
& $\mbox{\tt  nike air max} \cpc{CPC10} \mbox{Item}_2$ \\
 $\mbox{\tt  air max} \cpc{CPC11} \mbox{Item}_2$  & \\
\end{tabular}}

\vspace{0.5cm}

\noindent
The large erasers with an image size strictly greater than one are the following :\\

\noindent
{\footnotesize
\begin{tabular}{l|l}
large erasers & image\\
\hline
$\{\mbox{\tt nike}\}$ & $\mbox{\tt nike shoes}$, $\mbox{\tt nike soccer white}$, $\mbox{\tt  nike air max}$\\
$\{\mbox{\tt shoes}\}$ & $\mbox{\tt nike shoes}$,  $\mbox{\tt  adidas running shoes}$, $\mbox{\tt  superstar shoes}$\\
$\{\mbox{\tt large}\}$ & $\mbox{\tt large tee-shirt}$, $\mbox{\tt  large superstar shoes}$\\
$\{\mbox{\tt air}\}$ &  $\mbox{\tt  nike air max}$, $\mbox{\tt  air max}$\\
$\{\mbox{\tt max}\}$ &  $\mbox{\tt  nike air max}$, $\mbox{\tt  air max}$\\
$\{\mbox{\tt adidas}\}$ &  $\mbox{\tt  adidas running shoes}$,  $\mbox{\tt adidas superstar}$, $\mbox{\tt adidas superstar sneaker}$\\
$\{\mbox{\tt adidas, superstar}\}$ & $\mbox{\tt adidas superstar}$, $\mbox{\tt adidas superstar sneaker}$\\
$\{\mbox{\tt soccer}\}$ & $\mbox{\tt nike soccer white}$,  $\mbox{\tt soccer colored mens}$\\
$\{\mbox{\tt superstar}\}$ & $\mbox{\tt adidas superstar}$, $\mbox{\tt adidas superstar sneaker}$, $\mbox{\tt  large superstar shoes}$\\
\end{tabular}
}

\begin{figure}[htb!]
\centering 
\includegraphics[width=6.5cm]{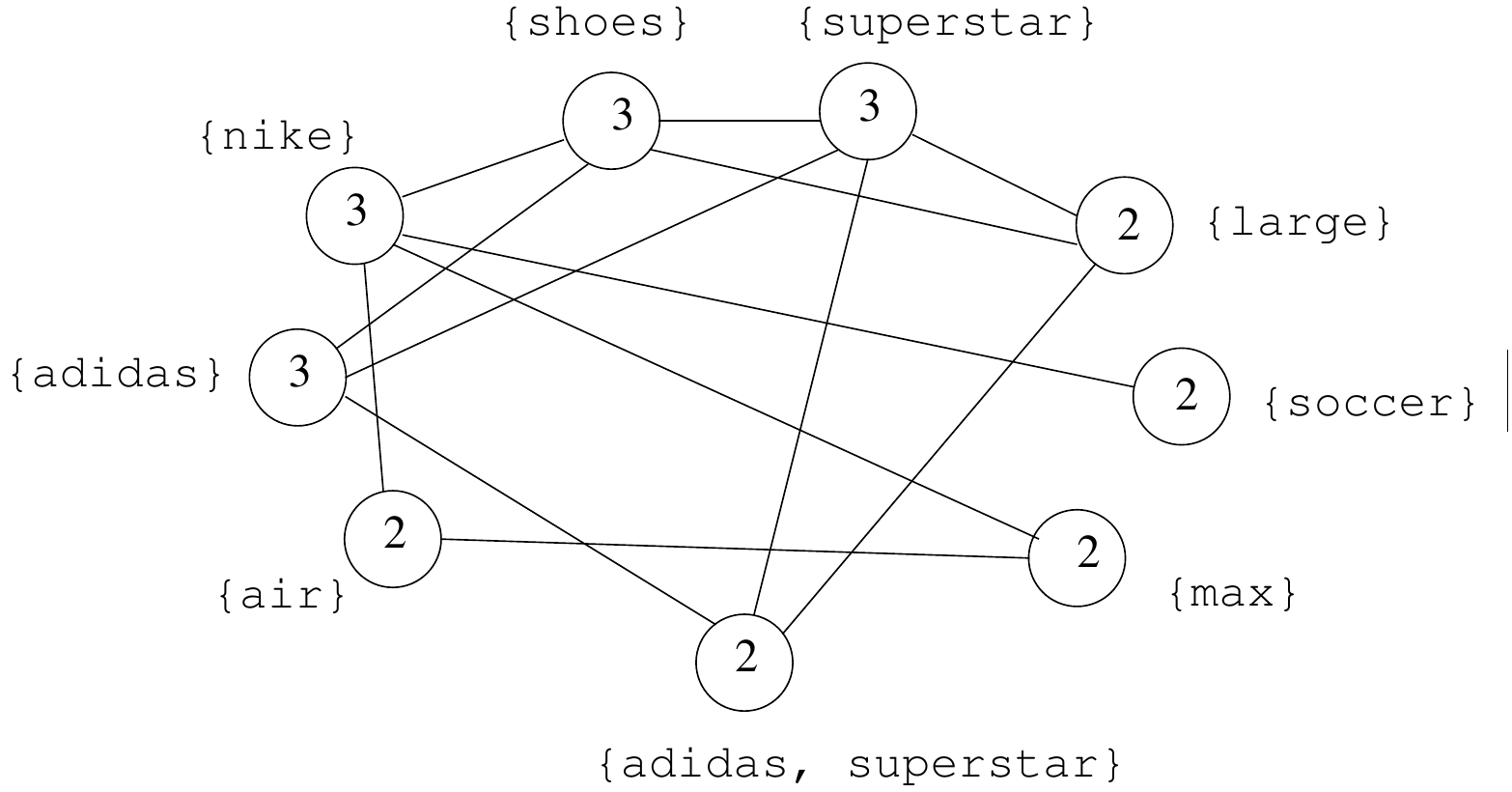} \quad \quad
\includegraphics[width=6.5cm]{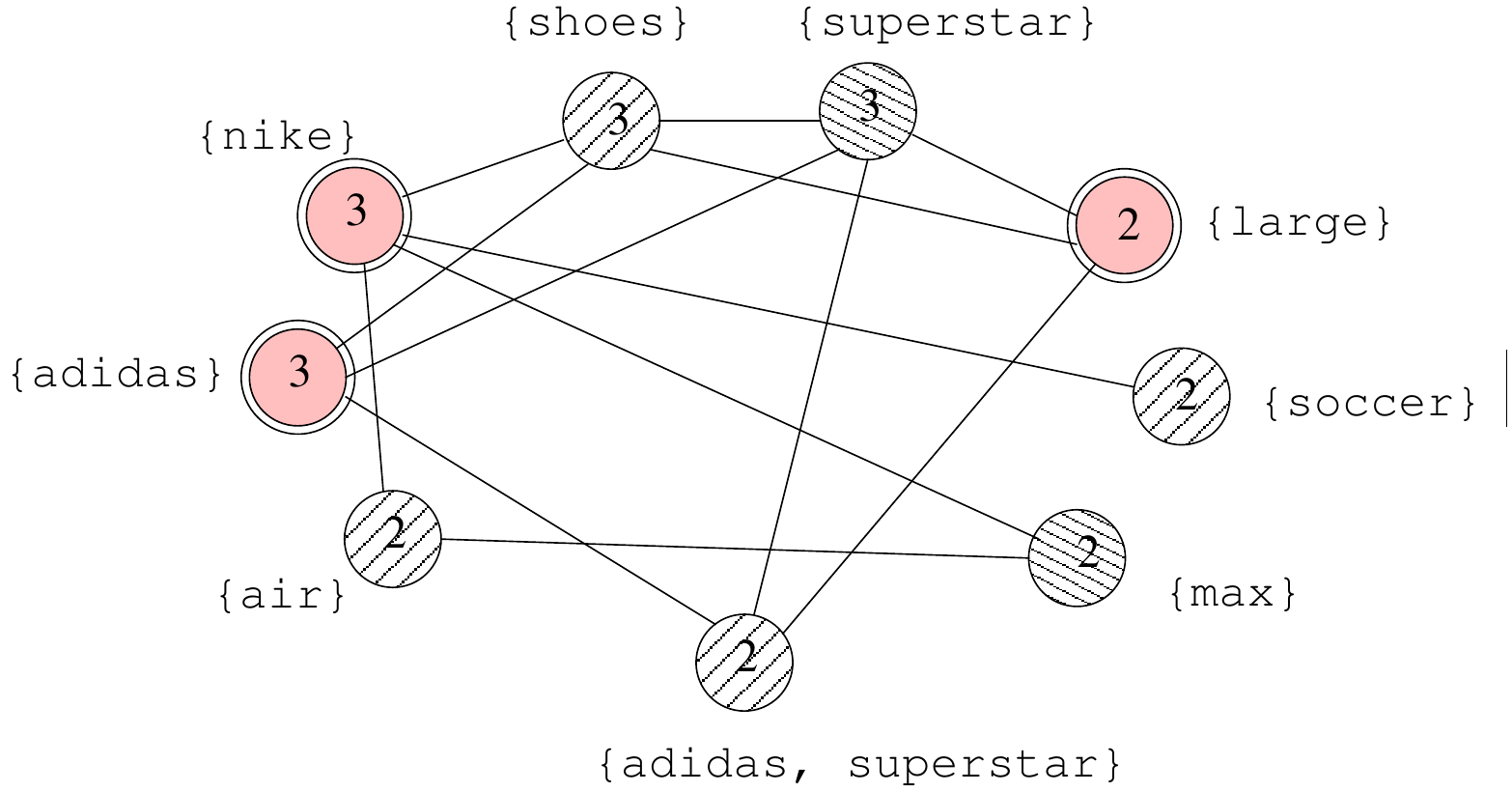}
\caption{Maximal weighted coloration example. The largest coloration
  found (nodes in double circles) touches $\frac{8}{11}$ keywords.}
 \label{coloring}
\vspace{-0.3cm}
\end{figure}

\vspace{0.5cm}
All image sizes of all large erasers are of size less than $\sqrt 11
\geq 3$, thus we keep all those erasers and build the graph $GE$,
given in figure \ref{coloring}.

We thus split $SK$ in 3 groups $sk_1 = $ $\{\mbox{\tt nike shoes}$,
$\mbox{\tt nike soccer white}$, $\mbox{\tt nike air max}\}$, $sk_2=$
$\{\mbox{\tt adidas running shoes}$, $\mbox{\tt adidas superstar}$,
$\mbox{\tt adidas superstar sneaker}\}$ and $sk_3=$ $\{\mbox{\tt large
  superstar shoes}$, $\mbox{\tt air max}$, $\mbox{\tt large
  tee-shirt}$, $\mbox{\tt garmin chronometer}\}.$

The erasers of $sk_1$, $sk_2$ and $sk_3$ are respectively $\{\mbox{\tt
  nike} (large)\}$, $\{\mbox{\tt adidas} (large)\}$ and $\{\mbox{\tt
  large} (large)$, $\mbox{\tt air max} (exact)$, $\mbox{\tt garmin
  chronometer} (exact) \}.$ We thus create 3 level 3 campaigns $C^1_3, C^2_3, C^3_3$ with
\begin{itemize}
\item $\mbox{\em Neg}(C^1_3)=$ $\{\mbox{\tt adidas} (large)$, $\mbox{\tt
  large} (large)$, $\mbox{\tt air max} (exact)$, $\mbox{\tt garmin
  chronometer} (exact) \}$
\item  $\mbox{\em Neg}(C^2_3)=$ $\{\mbox{\tt nike} (large)$, $\mbox{\tt
  large} (large)$, $\mbox{\tt air max} (exact)$, $\mbox{\tt garmin
  chronometer} (exact) \}$
\item $\mbox{\em Neg}(C^2_3)=$ $\{\mbox{\tt nike} (large)$, $\mbox{\tt
  adidas} (large) \}$\\
\end{itemize}

\noindent
We create 3 AdGroups for the first $C^1_3$ campaign: 
\begin{itemize}
\item  $\mbox{AdGroup}_1(C^1_3)$ corresponds to $\mbox{\tt nike
  shoes}$, and $\mbox{\em Neg}(\mbox{AdGroup}_1(C^1_3))=\{\mbox{\tt
  nike soccer white}$, $\mbox{\tt nike air max}\}.$
\item   $\mbox{AdGroup}_2(C^1_3)$ corresponds to $\mbox{\tt nike soccer white}$, and $\mbox{\em Neg}(\mbox{AdGroup}_2(C^1_3))=\{\mbox{\tt
  nike shoes}$, $\mbox{\tt nike air max}\}.$
\item   $\mbox{AdGroup}_3(C^1_3)$ corresponds to $\mbox{\tt nike air max}$, and $\mbox{\em Neg}(\mbox{AdGroup}_2(C^1_3))=\{\mbox{\tt
  nike shoes}$, $\mbox{\tt nike scoccer white}\}.$\\
\end{itemize}

\noindent
The remaining AdGroups of $C^2_3$ and   $C^3_3$ are built the same way.

\subsection{Adding A Rule (op1)}
\label{rule-op1}

Let us illustrate the procedure to add a rule. We present two examples, a simple and a more complex one.
First, let us add the rule
\begin{enumerate}
\item $\mbox{\tt nike jogging} \cpc{CPC12} \mbox{Item}_6$
\end{enumerate}

The keyword $\mbox{\tt nike jogging}$ is accepted only by $C^1_3$ and
erased by the others. This is the simplest case, we add $\mbox{\tt
  nike jogging}$ to $C^1_ 3$, create a new adgroup
$\mbox{AdGroup}_4(C^1_3)$ that corresponds to $\mbox{\tt nike
  jogging}$, and add the eraser $\mbox{\tt jogging}$ to the negative set of
all the others AdGroups of $C^1_3.$\\

\noindent
We now add the rule
\begin{enumerate}
\item $\mbox{\tt nike large shoes} \cpc{CPC13} \mbox{Item}_1$
\end{enumerate}

The keyword $\mbox{\tt nike large shoes}$ is erased by all low level
campaigns $C^1_3,$ $C^2_3,$ $C^3_3.$ Thus there are several
possibilities, as stated in Section \ref{adding}, depending of our
goal. If the goal is to minimize the number of adgroup changes, a
simple solution (point 2-(b) in Section \ref{adding}) is to create a
new campaign $C^4_3$ for the new keyword.  The negative keywords of $C^4_3$
have to stop all the other keywords excepted $\mbox{\tt nike large
  shoes}.$ A possibility is $\mbox{\em Neg}(C^4_3)=$ $\{\mbox{\tt nike
  shoes} (exact)$, $\mbox{\tt tee-shirt} (large)$, $\mbox{\tt garmin}
(large)$, $\mbox{\tt air} (large)$, $\mbox{\tt adidas} (large)$,
$\mbox{\tt soccer} (large)$, $\mbox{\tt superstar} (large) \}.$

\section{Perspectives}

We plan to implement this theoretical approach in a real shopping account and
then measure the volume modifications on which at this development state of our
technique we have only a restricted visibility. We are convinced that the notion of
keyword {\it eraser} and its associated algorithmics is just at its beginning.

\nocite{*}
\bibliographystyle{plain}

\end{document}